\documentclass[preprint,aps]{revtex4-1}

\usepackage{graphicx}
\usepackage{framed}
\usepackage{indentfirst}

\usepackage{dcolumn}% Align table columns on decimal point
\usepackage{bm}% bold math
\usepackage{color}% add textcolor
\usepackage{amsmath}% for subequations environment

\begin{document}

\title{ACER: A Framework on the Use of Mathematics in Upper-division Physics}

%\classification{01.30.Ib, 01.40.Di, 01.40.Fk,01.40.G-,01.40.gb}
%\keywords{physics education research, classical mechanics, math methods, task analysis, resources}

\author{Marcos D. \surname{Caballero}}
\author{Bethany R. \surname{Wilcox}}
\author{Steven J. \surname{Pollock}}
\affiliation{Department of Physics, University of Colorado Boulder, Boulder, CO 80309}	
%{
%  address={Department of Physics, University of Colorado, Boulder, CO 80309}
%  }

\author{Rachel E. \surname{Pepper}}
\affiliation{Departments of Integrative Biology and Civil \& Environmental Engineering, University of California, Berkeley, CA, 94720}

\begin{abstract}
At the University of Colorado Boulder, as part of our broader efforts to transform middle- and upper-division physics courses, we research students' difficulties with particular concepts, methods, and tools in classical mechanics, electromagnetism, and quantum mechanics. 
Unsurprisingly, a number of difficulties are related to students' use of mathematical tools (e.g., approximation methods). 
Previous work has documented a number of challenges that students must overcome to use mathematical tools fluently in introductory physics (e.g., mapping meaning onto mathematical symbols).  
We have developed a theoretical framework to facilitate connecting students' difficulties to challenges with specific mathematical and physical concepts. 
In this paper, we motivate the need for this framework and demonstrate its utility for both researchers and course instructors by applying it to frame results from interview data on students' use of Taylor approximations. 
\end{abstract}

\maketitle

\section{Introduction}
Each year 6,000 physics majors graduate from US colleges and universities after having completed rigorous coursework in upper-division physics \cite{mulvey2010focus}.
However, the PER community is accruing evidence that students throughout the major struggle with certain concepts, ideas, and tools \cite{pepper1289our,PhysRevSTPER.6.020115,Singh:2006tv,Smith:2010wx}.
These results are particularly troubling when considering the need to build on prior knowledge as our majors advance through the curriculum. 
Moreover, these persistent difficulties can make solving the long, complex problems in upper-division courses quite challenging.

%Part of the challenge in solving upper-division physics problems is the nature of the problems presented in these courses.
In particular, upper-division physics students solve many problems that require sophisticated physical ideas and mathematical tools (e.g., approximation methods). 
Students are taught these tools in their advanced mathematics courses and solve numerous abstract mathematical exercises.
Yet, students still struggle to employ these tools in their physics courses \cite{2012AIPC.1413...89W}.
In physics, mathematical tools serve a different purpose; they are used to make inferences about physical systems.
Furthermore, students must synthesize additional knowledge (e.g., conceptual physics knowledge) to apply mathematical tools to physics problems \cite{Redish:2009un}.

In contrast to the substantial work addressing problem solving in introductory physics courses \cite{Hsu:2004kh}, less work has been done in upper-division physics \cite{ambrose2004investigating,chasteen2008transforming,PhysRevSTPER.7.010101}.
Prior upper-division research has focused on noting and cataloging conceptual difficulties.
As students grapple with longer and more complex problems, it is increasingly important to integrate research on students' conceptual knowledge with research on their use (or misuse) of mathematical tools.
Such integration will provide a more complete understanding of how students in the upper-division solve problems.

In this paper, we begin to make inroads into a synthesis of conceptual knowledge and mathematical tool usage from a theoretical perspective.
Previous work in introductory physics has produced several helpful theoretical frameworks that serve a variety of purposes: coordinating multiple theories of learning \cite{Tuminaro:2004vr}, building on lessons from mathematics education \cite{rebello:2011ad}, and providing a logical construction for solving problems \cite{heller1995competent}.
We built upon this foundation to develop a framework to address how mathematical resources are Activated, Constructed, Executed, and Reflected upon (ACER).
The ACER framework was designed to aid both instructors and researchers in exploring when and how students employ particular mathematical tools when solving canonical exercises from upper-division physics courses.
We have also found the framework useful when developing new problems, and critiquing old ones.
This paper discusses the design of this framework, demonstrates its utility with a particular example from middle-division classical mechanics
(Taylor approximations), and closes with a discussion of implications and future investigations.  

\vspace*{-14pt}
\section{The ACER framework}

To help organize our observations of students' problem solving difficulties in upper-division physics courses in terms of students' conceptual knowledge and their use of mathematical tools, we have developed a theoretical framework that applies to the types of complex problems students encounter in these courses.
The framework is grounded in task analysis, a method developed to uncover the tacit knowledge used by experts to perform complex tasks \cite{catrambone1996generalizing,Catrambone:vf}, and resource theory, a model of the nature of knowledge and how it is activated and employed \cite{Rumelhart:1980wp,Hammer:2005th}.
Development of the framework was motivated by observing common difficulties in student solutions to Taylor approximation problems.
We performed a task analysis on a number of these problems, which required reflecting on, documenting, and organizing the elements necessary to complete each problem.
After several iterations, we organized the various elements into components that highlight the physical and mathematical concepts being activated and employed while solving Taylor approximation problems.

\begin{figure}
\includegraphics[clip, trim=12mm 50mm 12mm 14mm, width=0.90\linewidth]{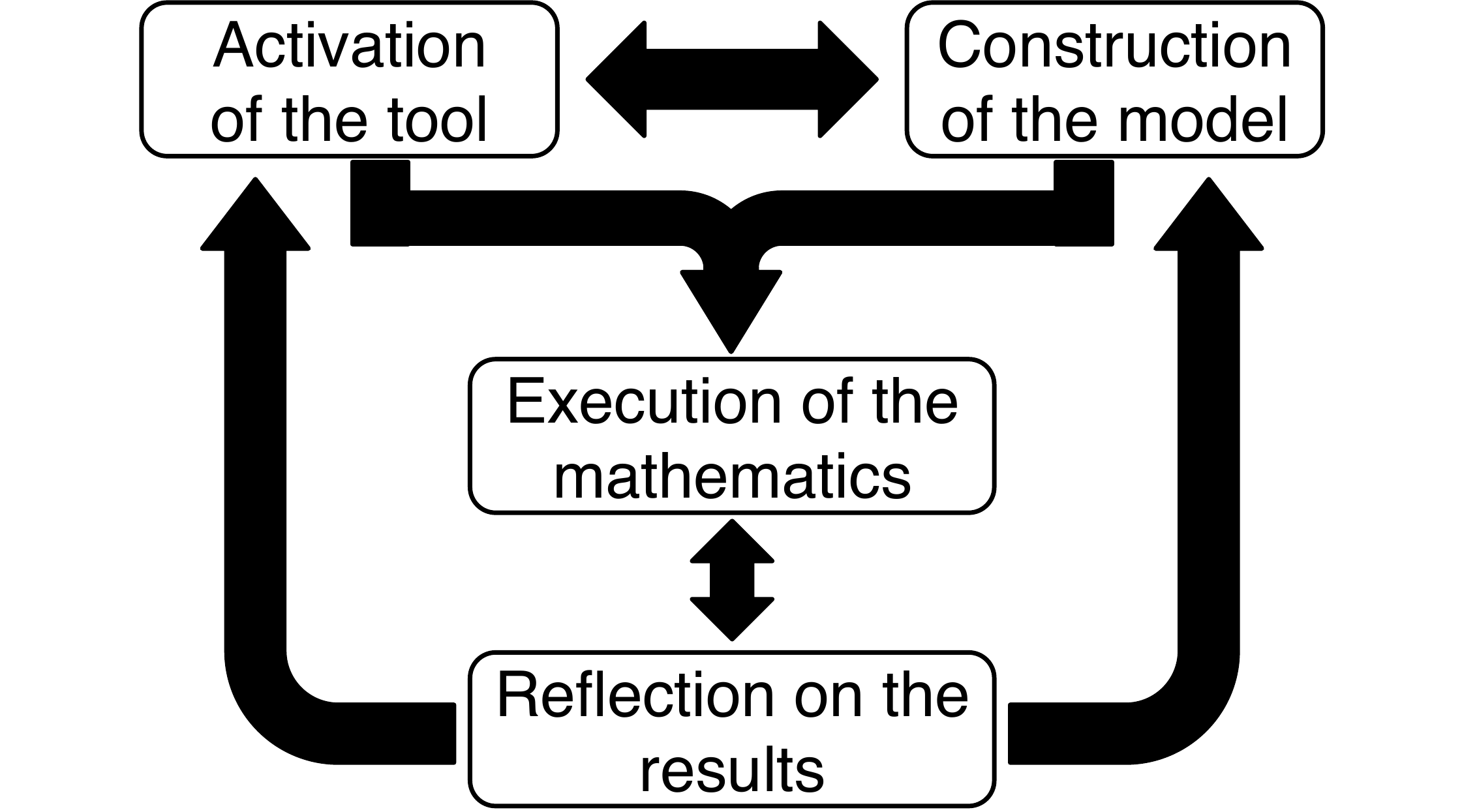}
\caption{A visual representation of the ACER framework. %This framework provides a common language and convenient scaffold for investigating student difficulties with employing mathematical tools in physics.
}\label{fig:acer}
\end{figure}

Our framework is organized around 4 components: {\it Activation of the tool}, {\it Construction of the model}, {\it Execution of the mathematics}, and {\it Reflection on the results} (ACER).
ACER frames the challenges that students are likely to encounter in their coursework (i.e., solving ``back of the book'' style problems).
To solve such a problem, one must determine which mathematical tool is appropriate for the model of the physical system they have constructed.
Then, a series of mathematical steps are executed that facilitate the development of a solution, which must be checked for errors and compared against established or known results. 
A convenient way to visualize the ACER framework is shown in Fig. \ref{fig:acer}.
In Fig. \ref{fig:acer}, we are not suggesting that all physics problems are solved in some clearly organized fashion, but a well articulated, complete solution involves all components of the ACER framework.
The ACER framework is not general enough to be applied to experimental or open-ended problems.
However, its targeted focus means that it can be operationalized for a variety of common mathematical tools (e.g., direct integration of distributed charge \cite{wilcox:2012a}).
Below, we describe the details of each component in the context of Taylor approximations.

{\bf Activation of the tool:} 
A problem statement contains a number of explicit or implicit cues that might activate any of a number of resources (or resource networks) associated with one or more mathematical tools \cite{Hammer:2005th}.
Each student has their own particular association between cues and resources.
%;Particular cues can activate different resources depending on what a student associates with those cues.
Some resources that are activated might help complete the problem and others might misdirect students' efforts. 
For the problem shown in Fig. \ref{fig:grav},  these cues include: the goal of the problem (``determine an \emph{approximate} expression for the difference in the gravitational potential''), as well as language and symbols that suggest some quantity ($d$) is much smaller than some other quantity ($R$).
These cues are intended to
%(e.g., ``determine an approximate expression'', language suggesting a comparison, and vastly different length scales) 
activate resources associated with Taylor approximation.
%In another example, a gravitational problem might be solved by using Gauss' law or through direct integration.
%It is the activation of a particular resource (e.g., spherical symmetry) that might lead one to employ Gauss' law instead of direct integration.

\begin{figure}
\begin{minipage}{1\linewidth}
\begin{framed}
\vspace*{-8pt}
{ An astronaut is in orbit around the Earth at a distance of $R$ from the center of the Earth. Another astronaut is in a closer orbit ($R - d$). The difference in the strength of the gravitational potential between the astronauts is, $\Delta \phi = \frac{G\;M_E}{R}-\frac{G\;M_E}{R-d}$.
Determine an approximate expression for the difference in the gravitational potential when the astronauts are very near each other.}
\vspace*{-8pt}
\end{framed}
\end{minipage}
\caption{A sample problem with embedded cues that activate certain resources associated with Taylor approximation.}\label{fig:grav}
\end{figure}

{\bf Construction of the model:} 
In physics, mathematics represents a simplified picture (i.e., a model) of a real system where each symbol has a particular physical meaning \cite{Redish:2009un}.
A mathematical model is typically needed to develop a solution to a physics problem.
%We map physical meaning onto the symbols in these mathematical models .
%Problems are questions about the behavior of systems and models predict that behavior.
%Which symbols are constants? 
%Which symbols are variables or parameters? 
%Which variables or parameters are typically of interest? 
%What are the units? 
Mathematical models used in physics are typically written in a compact form (e.g., $\phi = - \int G\,dM/r$) and the identity of variables and parameters must be known or discovered.
Given a specific physical situation, the use of different representations (e.g., diagrammatic or graphical) to construct elements precedes the expression of mathematical model. 
In some cases, a mathematical model might be provided but requires meaning be mapped onto the expression.
For example, the equation given in Fig. \ref{fig:grav} was constructed by an instructor and was provided to students, but additional work is needed to understand the model (e.g., recognizing the small expansion parameter is $d/R$).
%gravitational problems involving distributed mass require choosing an appropriate coordinate system, expressing the differential mass element in an appropriate form, determining limits of integration, and expressing the difference vector in the appropriate coordinates.

{\bf Execution of the mathematics:} 
Transforming the math structures (e.g., unevaluated integrals) in the construction component into relevant mathematical expressions (e.g., evaluated integrals) is often necessary to uncover solutions \cite{Redish:2009un}.
%Moreover, experts know the rules for valid use of the tool.
Each mathematical tool requires a specific set of steps and basic knowledge.
For example, executing a Taylor approximation may require knowledge of common expansion templates (e.g., $\sin x \approx x + x^3/3! + \dots$) and how to adapt these templates to the mathematical model developed previously.
Alternatively, one might need to know how to compute derivatives of complex functions.
The mathematical procedures performed in this component are not, at least to experts, context free.
In addition to employing base mathematical skills, experts maintain awareness of the meaning of each symbol in the expression (e.g., which symbols are constants when taking derivatives).

{\bf Reflection on the results:} 
Expressions that are developed in upper-division physics courses are not superficial manipulations of mathematical expressions from textbooks or notes.
These expressions are new entities that have predictive power and can provide greater additional insight into the behavior of the system.
Reflecting on derived expressions is crucial to provide confidence in their predictions and insights (e.g., how can we  know a particular expression is the correct one?).
At the most basic level, reflection involves checking expressions for errors (e.g., by checking their units). 
Comparing the predictions 
%of these expressions
to established or known results (e.g., determining its limiting behavior) is also necessary to gain confidence in these expressions.
If a mistake occurred in executing the mathematics or, perhaps, some incorrect element was used in constructing the model, reflecting on the result in various ways can help uncover these errors.

The ACER framework is most closely related to Redish's ``Use of mathematics in physics'' \cite{Redish:2009un} and the ``Logical Problem Solving Strategy'' of Heller, et al. \cite{heller1995competent}; but, we distinguish it from both in its intent, its focus, and its utility.
Our framework was not intended to be a model for student reasoning nor to provide a series of steps for solving problems. 
It provides a scaffold onto which elements of a student's solution can be organized by researchers or instructors.
In doing so, ACER can help describe where students are being challenged (e.g., students produce nonsensical solutions), and can provide reasons why these difficulties exist (e.g., problems or activities focus on {\it Execution} while neglecting {\it Reflection}).

\vspace*{-14pt}
 \section{Employing ACER -- Taylor Series}
We performed eight video-taped think-aloud interviews to investigate students' use of Taylor approximations. 
%; one series of interviews occurred before the development of the framework and the other after.
The eight participants were physics, engineering physics, and astrophysics majors recruited from the first (6 participants) and second (2 participants) semester classical mechanics courses at CU Boulder.
Taylor series had been covered in the first semester course several weeks prior to the first study.
%Six participants were enrolled in the first semester course and two in the second semester course.
Participants tended to be the more motivated students in the course, but their exam scores reflected the full gamut of passing grades (A to D).

\begin{figure}
\begin{minipage}{1\linewidth}
\begin{framed}
\vspace*{-8pt}
{A small sphere (mass, m) is free to slide inside a frictionless cylinder of radius $R$. If placed at the equilibrium point, $\phi = 0$ (shown below), the ball does not move.}

\begin{minipage}{0.35 \linewidth}
\begin{center}
\includegraphics[clip,trim=72mm 50mm 75mm 15mm, width=0.90\linewidth]{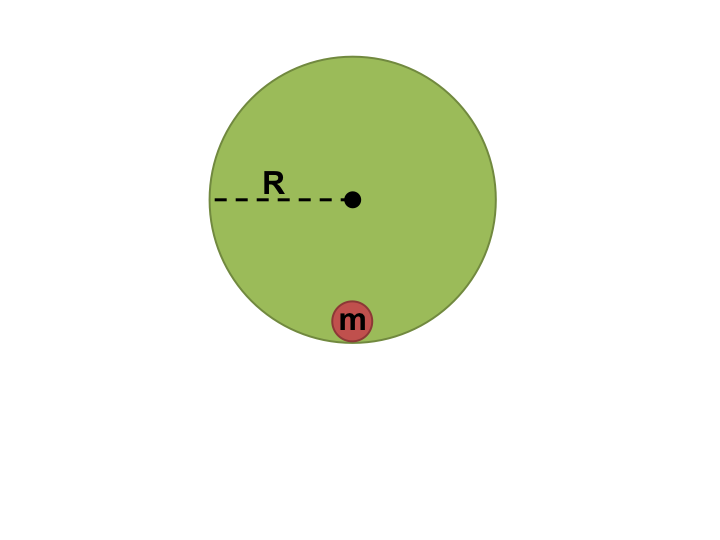}
\end{center} 
 \end{minipage}
\begin{minipage}{0.60 \linewidth}
{At other non-equilibrium angles ($\phi$), the gravitational potential energy for this system is given by $U(\phi) = mgR (1 - \cos \phi)$. Find an approximate expression for the gravitational potential energy for $\phi$ near $\phi = 0$.}
 \end{minipage}
\vspace*{-16pt}
 \end{framed}
\end{minipage}
\caption{A context-laden Taylor approximation problem with implicit cueing used in think-aloud interviews.}\label{fig:cyl}
\end{figure}

Both studies asked students to solve a series of Taylor approximation problems.
Students' written solutions were captured using a smartpen with an embedded audio recording device (Livescribe pen).
Problems included both formal math and context-laden physics questions (e.g., Fig. \ref{fig:cyl}). 
In the first study, formal math questions were asked first, and context-laden physics questions with explicit cueing (e.g.,``perform a Taylor expansion'') were asked later.
We developed and, later, applied the ACER framework to organize observations from the first study.
ACER demonstrated that the first study limited the possibility of observing attempts to process implicit cues.
%Changes to the study design which included the use of implicit cueing were informed by the development of the ACER framework.
The second study began with context-laden physics questions with implicit cueing (e.g., ``find an approximate expression''); formal math questions were delayed to the end.
%Questions required students to manipulate mathematical models to develop approximate expressions.
Students were asked to describe how they constructed their approximate formulae
% (e.g., determining the small parameter).
%Questions in both studies asked for descriptions of the expansion parameter (e.g., $x-x_0$) including direct questioning about how the expansion variable (e.g., $x$) was determined.
and to reflect on the physical meaning of the terms in their expressions by comparing them to known results.
% (e.g., time of flight in the absence drag).
Video data was analyzed by identifying each key element of the framework that appeared in the students' solutions.
Gestures were used to identify which parts of the solutions were being discussed in the interview.
With the data we collected in these studies, we have started to organize the challenges students face with Taylor approximations using the ACER framework.

{\bf Activation of the tool:}
Some upper-division physics students are just beginning to learn the ``language of physics'' and have not yet internalized the implicit cues that activate the use of approximation methods.
No participant in the explicit cueing study failed to start a problem with a Taylor approximation. 
However, when solving problems with implicit cueing, 2 of 4 participants initially plugged in the given numeric value (e.g., $\phi$ = 0 in Fig. \ref{fig:cyl}) to determine the approximate expression (e.g., $U(\phi) \approx 0$).
%After additional cueing from the interviewer, one of these participants attempted to use a geometric argument (i.e., use $\phi$ to find the ``approximate'' height above ground in Fig. \ref{fig:cyl}) to determine the approximate expression.
After these 2 participants began working the formal math problems, both asked to return to the previous context-laden problems to include approximation methods in their answers.

{\bf Construction of the model:} 
Our studies did not deeply investigate model construction; students were often given formulae from which to work.
From this limited investigation, we found that students are able to map meaning onto symbols, but they struggle to identify the relative scales of relevant variables in the problem (e.g., $d \ll R$ in Fig. \ref{fig:grav}).
%However, to utilize a Taylor expansion, a student must determine for which variable the expansion should be carried out and, in doing so, identify the relative scales of relevant variables in the problem (e.g., $d \ll R$ in Fig. \ref{fig:grav}).
In both studies, participants appropriately focused their attention on variables relevant to the Taylor approximation (e.g., $\phi$ in Fig. \ref{fig:cyl}) rather than constants (e.g., $g$) and parameters (e.g., $R$).
Furthermore, no participant had significant difficulties interpreting the provided mathematical expressions.%; 
% given in the problem statements; 
%most of the expressions were likely familiar to the participants.
However, 
%in both studies, 
%participants were unable to satisfactorily discuss the relative scales of variables. 
seven participants claimed their various expansions provided a ``good'' approximation to the original expressions if the variable %(e.g., $\phi$) 
was ``small compared to 1''
%Similar statements were made 
regardless of the expression under consideration or the presence of a natural comparative scale.

%For example, when determining the approximate expression for velocity of an object in the presence of drag for ``short time,'' students identified that in this case the time should be less (or much less) than 1.
%The natural time scale for this problem is $1/\beta$; a concept that is covered both in lecture and on homework.
%We also have preliminary evidence from exam questions that indicates students struggle to construct the appropriate formula when the expansion point (e.g., $x_0$) is non-zero.

{\bf Execution of the mathematics:} 
When computing the Taylor approximation of a function, most students elect to use the formal definition of Taylor series (Eq. \ref{eqn:taylor}).
% over optimized methods.
%Calculating the Taylor expansion of a function, for most participants, was in an exercise in recalling the generalized formula and computing derivatives.
%(Eq. \ref{eqn:taylor}) and computing derivatives.
% \begin{equation}
% f(x) = \sum_{n=0}^{\infty}\frac{f^{(n)}(x_0)}{n!}\left(x-x_0\right)^{n}
% \label{eqn:taylor}
% \end{equation}
When requested, formulae for Taylor expansions of common functions were given to participants, but, in most instances, participants recalled or requested the formal definition.
Using the formal definition was not incorrect, but led to a variety of mathematical mistakes such as taking derivatives incorrectly and forming non-polynomial expansions.
%,
\vspace*{-8pt}
\begin{equation}
f(x)=\sum_{n=0}^{\infty} \frac{f^{(n)}(x_0)}{n!} (x-x_0)^n.
\label{eqn:taylor}
\end{equation}
\vspace*{-8pt}

All eight students solved the formal math problems correctly using the correct reasoning.
%Moreover, most students had no trouble grinding through the machinery of a Taylor expansion once it was activated. 
%Although, students often used the formal definition of a Taylor expansion; o
Most students took derivatives of the equation given in Fig. \ref{fig:cyl}.
Only 1 student employed a common optimized method: mapping the given expression onto a certain known expansion template (e.g., $\cos(x)\approx1-x^2/2!$).
%Using the formal definition is not incorrect, but it provides additional opportunities to make algebraic and calculus mistakes.

{\bf Reflection on the results:} 
When prompted, most students reflect on newly constructed expressions without a clear purpose.
%Some participants discovered minor errors in their expressions, but m
Most were unable to connect these expressions back to the physics under investigation.
%All of the correct expressions in our Taylor series studies were connected to well known results from less rich physical model (e.g., projectile motion without drag).
In each context-laden physics problem, participants were asked to discuss their approximate expressions and reflect on its predictions.
We aimed for participants to connect these expressions back to commonly understood phenomenon (e.g., projectile motion without drag).
For several participants, these ``forced'' reflections helped uncover minor mathematical mistakes.
% in some problems.
Typically, this occurred when a participant checked the units of various terms.
However, only 1 student discussed the connection between individual terms and the relevant physics (e.g., ``That looks like [the potential of] a harmonic oscillator'', gesturing to the approximate expression).
The other seven students discussed terms superficially (``that's the drag term'') or not at all (``yeah, that looks different from the usual equation'').
 
\vspace*{-14pt}
\section{Concluding Remarks}\label{sec:closing}
We are using the ACER framework to help both instructors and researchers explore when and how students employ Taylor approximations when solving typical exercises from upper-division physics courses.
%Our preliminary studies suggest that students have not yet internalized the cues to activate the use of approximation methods. 
The ACER framework helps to untangle students' sophisticated mathematical and physics difficulties and provides a convenient scaffold on which to hang these challenges.
In a sense, the ACER framework addresses many important elements that define what it means to use mathematics in upper-division physics well.
As such, it also provides a means to critique old and design new problems.
Our investigations demonstrate that current instruction fails to enculturate students to the implicit cues that activate the use of approximation methods.
Moreover, working problems with a deeper emphasis on identifying small parameters and mapping known expansion templates to mathematical models in a variety of contexts would likely benefit many students.
Finally, when prompted, most students reflect on newly constructed expressions superficially, at most, checking the units of their expression. 
Meaningful reflections are important for connecting the math that was performed and the physics it describes.
Instruction should highlight the need to gain confidence in the predictions of and insights gained from new expressions.

%In a sense, the ACER framework provides an ambitious definition of what it means to use mathematics in upper-division physics.
%As such, it provides a scaffold for critiquing old and designing new problems.
%For example, physics students are rarely asked to choose which mathematical tool to employ to solve a given problem.
%%When and how do students learn to employ approximation methods such as Taylor expansions?
%Direct wording in the problem statement or classroom instructional practices tend to provide explicit cues.
%Alternative problems might be developed to highlight the importance of language and symbols in a student's decision to use a Taylor approximation.
 
The ACER framework is under continual refinement.
At present, where and how other important activities like coordinating representations, interpretation and prediction, and metacognition fit is an open research agenda. 
However, the ACER framework has already proven useful in guiding future research efforts. 
In future Taylor series studies, we plan to unpack the complexities of identifying small parameters (Construction) and gaining confidence in expressions (Reflection).
In addition, our framework is being used in junior-level electromagnetism to explore difficulties with direct integration of charge distributions \cite{wilcox:2012a}.
Future work will be expanded to include separation of variables in boundary value problems, and the use of direct integration and Gauss' law in gravitational problems.  
These studies of a variety of mathematical tools will help to further refine the ACER framework.

%The ACER framework is being used to guide future research efforts. 

%In addition, we have begun to use the ACER framework in junior-level electromagnetism to frame student difficulties with direct integration of charge distributions.
%Future work will be expanded to include separation of variables in boundary value problems, the use direct integration and Gauss' law in gravitational problems, and the use of certain techniques from complex analysis in electromagnetism.  
%Finally, longitudinal studies are planned to uncover when crucial practices, like Reflection, are ``picked up'' by physics majors, if at all.

%%%%%%%%%%%%%%%%%%%%%%%%%%%%%%%%%%%%%%%%%%%%%%%%
%% BACKMATTER
%%%%%%%%%%%%%%%%%%%%%%%%%%%%%%%%%%%%%%%%%%%%%%%%

\vspace*{-14pt}
\section{Acknowledgments}
We gratefully acknowledge the contributions of the student participants. 
Particular thanks to the members of the PER@C group and the three anonymous reviewers for their thoughtful insights.
This work is supported by University of Colorado's Science Education Initiative.

\bibliographystyle{aipproc}   % if natbib is available

\vspace*{-10pt}
\bibliography{math_phys}

\IfFileExists{\jobname.bbl}{}
 {\typeout{}
  \typeout{******************************************}
  \typeout{** Please run "bibtex \jobname" to optain}
  \typeout{** the bibliography and then re-run LaTeX}
  \typeout{** twice to fix the references!}
  \typeout{******************************************}
  \typeout{}
 }

\end{document}